\RequirePackage{lineno}
\documentclass[prl,preprintnumbers,amsmath,amssymb,superscriptaddress,showpacs, twocolumn]{revtex4-1}
\usepackage[utf8]{inputenc}

\usepackage{hyperref}

\usepackage{graphicx}
\usepackage{upgreek}
\usepackage{color}
\usepackage[separate-uncertainty = true]{siunitx}

\begin{document}

\title{Achiral tilted domain walls in perpendicularly magnetized nanowires}

\author{B. Boehm}
\email{boe@zurich.ibm.com}
\affiliation{IBM Research - Zurich, 8803 R\"uschlikon, Switzerland}

\author{A. Bisig}
\affiliation{IBM Research - Zurich, 8803 R\"uschlikon, Switzerland}

\author{A. Bischof}
\affiliation{IBM Research - Zurich, 8803 R\"uschlikon, Switzerland}

\author{G. Stefanou}
\affiliation{School of Physics and Astronomy, University of Leeds, Leeds LS2 9JT, United Kingdom}

\author{B. J. Hickey}
\affiliation{School of Physics and Astronomy, University of Leeds, Leeds LS2 9JT, United Kingdom}

\author{R. Allenspach}
\email{ral@zurich.ibm.com}
\affiliation{IBM Research - Zurich, 8803 R\"uschlikon, Switzerland}
\date{April 7th, 2017}

--------------------------

\pacs{}



\begin{abstract}
Perpendicularly magnetized nanowires exhibit distinct domain wall types depending on the geometry. Wide wires contain Bloch walls, narrow wires N\'eel walls. Here, the transition region is investigated by direct imaging of the wall structure using high-resolution spin-polarized scanning electron microscopy. An 
achiral
intermediate wall type is discovered that 
is unpredicted by established theoretical models. With the help of micromagnetic simulations, the formation of this novel wall type is explained.
 
\end{abstract}

\maketitle

In recent years, domain walls in perpendicularly magnetized materials have been intensely investigated because they are narrower than in in-plane systems and therefore, when used to store a data bit, promise higher storage density. In perpendicularly magnetized systems, domain walls are of Bloch type, i.e., the magnetization rotates within the wall plane. In in-plane magnetized systems, in contrast, diverse wall types exist. In bulk, again Bloch walls prevail, whereas in thin films, the energetically favored wall type is a N\'eel wall, i.e., the magnetization rotates perpendicularly to the plane of the wall. In between, a finite film-thickness range exists in which domain walls are neither of Bloch nor of N\'eel type. They are characterized by more complex arrangements of spins, such as zigzag patterns \cite{shtrikman_1960}, cross-ties \cite{torok_1965} or continuous asymmetric deformations \cite{rivkin_2009}. The Bloch wall is the energetically preferred state in perpendicularly magnetized films irrespective of film thickness. N\'eel walls can be made the ground state by changing the geometry to wires \cite{martinez_2011,dejong_2015} or adding constrictions \cite{jubert_2005}, by applying magnetic fields \cite{miron_2011, ryu_2013}, or by introducing Dzyaloshinskii--Moriya exchange interaction (DMI) \cite{thiaville_2012}.

The Bloch--N\'eel wall transition in perpendicularly magnetized nanowires, as a function of the wire width, was indirectly observed by measuring a change in the anisotropic magnetoresistance (AMR) \cite{koyama_2011}. Most recently, it was studied analytically \cite{dejong_2015}. A direct observation of this transition in real space is missing. Both Bloch and N\'eel walls were observed in thin films by spin-polarized scanning tunneling microscopy \cite{meckler_2009}, in multilayers by spin-polarized low-energy electron microscopy \cite{chen_2013}, and in nanowires by optically monitoring the Zeeman shift of the electron spin in a nitrogen-vacancy defect in diamond \cite{tetienne_2015}.

\begin{figure}[h!]
  \begin{center}
  \leavevmode
  \includegraphics[width=0.5\textwidth]{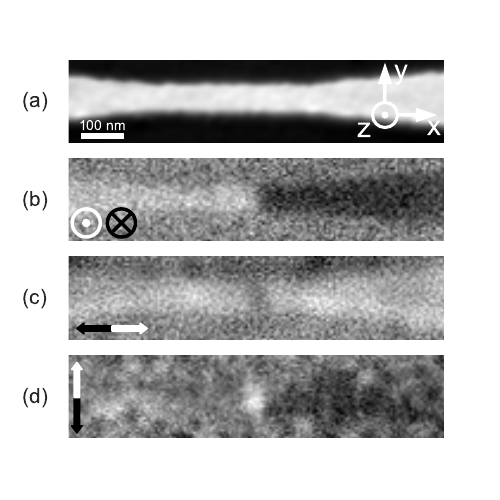}
 \end{center}
  \vspace*{-5mm}
	\caption{(a) Scanning electron micrograph of a 70-nm-wide ferromagnetic ``bowtie'' nanowire. The spin polarization micrographs show the magnetization component along the (b) \textit{z}-, (c) \textit{x}-, (d) \textit{y}-direction. The black/white contrast shows the direction of the magnetization, as indicated by the arrows. A domain wall is found at the narrowest part of the wire, where also the wire width of \SI{70}{\nano\meter} is measured. The wall has magnetization components in both in-plane directions. In addition, a slight canting of the domain wall is observed. 
	The images (c) and (d) were Gaussian filtered to improve visibility.	
	}
	\label{fig1}
\end{figure}

In this work, we investigate domain walls at the Bloch--N\'eel wall transition in flat nanowires (or ``nanostrips'') with perpendicular magnetic anisotropy as a function of the nanowire width. We image the wall structure in real space using high-resolution spin-polarized scanning electron microscopy (spin-SEM), which is capable of determining the specimen's magnetization by measuring the spin polarization of the low-energy secondary electrons emitted. We find Bloch domain walls in wide nanowires and N\'eel walls in narrow wires, and an intermediate domain wall type in between. This intermediate wall is characterized by a tilted magnetization direction pointing neither along the wire nor perpendicular to it. It marks the Bloch--N\'eel wall transition and has never been imaged before. We determine the domain wall profile and the azimuthal angle of the magnetization by fitting the profiles with a one-dimensional (1D) wall model. Micromagnetic simulations confirm this continuous transition via an intermediate domain wall and prove that this wall requires a nontrivial 2D arrangement of the spins.


The magnetic thin films Ta(\SI{5}{\nano\meter})/Pt(\SI{3}{\nano\meter})/ $\left[{\rm Co}(\SI{0.4}{\nano\meter})/{\rm Ni}(\SI{0.7}{\nano\meter})\right]_{3}$/Co(\SI{0.55}{\nano\meter})/Pt(\SI{1.5}{\nano\meter}) were sputter deposited onto a Si/SiO$_x$(\SI{6}{\nano\meter}) substrate. Vibrating sample magnetometry was used to determine the saturation magnetization $M_{\rm S}= \SI{5.7 \pm 0.2 e5}{\ampere\per\meter}$ and the perpendicular anisotropy $K_{\rm u}=\SI{2.5 \pm 0.4 e5}{\joule\per\cubic\meter}$ of the films. The nanowires were fabricated by Ar-ion milling through a 15-nm-thick Al mask that was patterned by electron-beam lithography and lift-off. The wires were shaped like a narrow bowtie, thus trapping the domain wall close to the center of the constriction upon application of an alternating perpendicular field to inject domain walls from adjacent large pads. Prior to magnetic imaging, the samples were sputtered with a ${\rm Xe}^{+}$ ion beam of \SI{1}{\kilo\volt} energy at normal incidence and a beam dose of $\approx\SI{1.7}{\coulomb\per\square\meter}$ in our ultra-high-vacuum system (\SI{1 e-10}{\milli\bar}) to remove 1 nm of the Pt capping layer, thereby enhancing the spin polarization of the secondary electrons while still keeping the perpendicular anisotropy of the Co/Pt interface. The sputtering process is controlled by monitoring the atomic composition of the thin-film surface by Auger electron spectroscopy.

Magnetic nanowires of different widths were imaged in our spin-SEM \cite{allenspach_2000}. We present results from nanostrips with width between \SI{57}{\nano\meter} and \SI{300}{\nano\meter}, focusing mainly on a 70-nm-wide wire. The wire width was determined at the position of the domain wall directly from the scanning electron micrograph with an uncertainty of \SI{5}{\nano\meter}. The spin detector measures two components of the magnetization simultaneously. All three magnetization components were accessed by taking two consecutive images, the second one after rotating the sample by \SI{90}{\degree}.

\begin{figure}
  \begin{center}
  \leavevmode
  \includegraphics[width=0.48\textwidth]{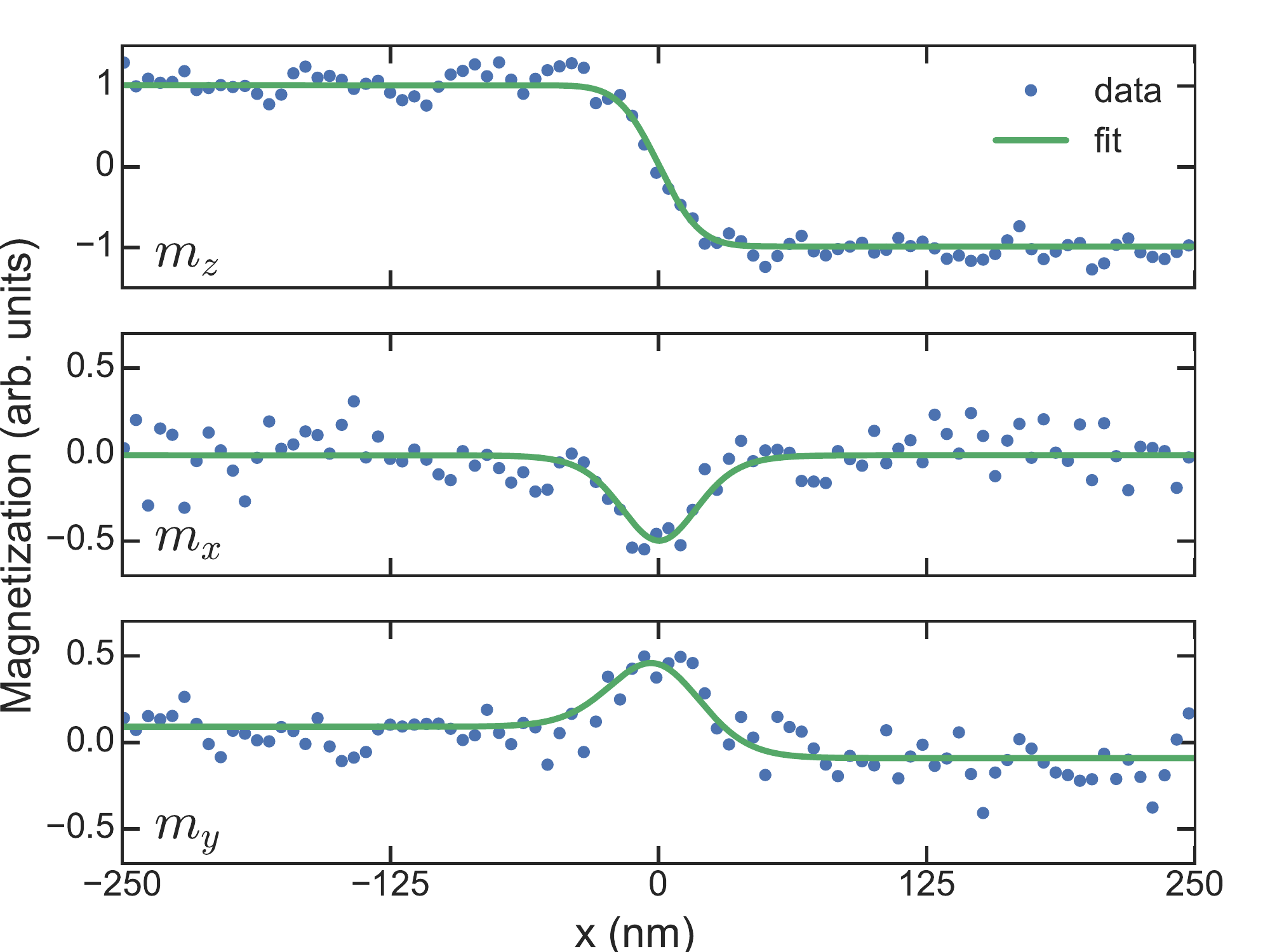}
 \end{center}
  \vspace*{-5mm}
	\caption{The magnetization profiles $m_i(x)$ ($i = x,y,z$) of the 70-nm-wide nanowire as a function of the distance from the center of the wall. The perpendicular and in-plane magnetization components were fitted with the 1D domain wall model and considering the finite resolution of the microscope.
			}
	\label{fig2}
\end{figure}

Figure \ref{fig1} shows that the domain wall in the 70-nm-wide wire has magnetization components in both in-plane directions: It is neither a pure Bloch nor N\'eel wall, but intermediate between the two. A closer look reveals that the domain wall is inclined by \SI{11 \pm 6}{\degree} with respect to the wire's cross section. In Fig. \ref{fig2} we plot the domain wall profiles $m_{\rm i}(x)$ ($i = x,y,z$) averaged along the \textit{y}-direction across the wire. The wall profile in the 1D model is \cite{jung_2008}:
\begin{align}\begin{split}
m_{x}(x) & =  \cos(\psi) / \cosh\left(\frac{x-x_0}{\lambda}\right), \\
m_{y}(x) & =  \sin(\psi) / \cosh\left(\frac{x-x_0}{\lambda}\right), \\
m_{z}(x) & =  -\tanh\left(\frac{x-x_0}{\lambda}\right), 
\end{split}\end{align}
where $\lambda$ is the domain wall width and $x_0$ is the center of the wall. The azimuthal angle $\psi$ is \SI{0}{\degree} (or \SI{180}{\degree}) for a N\'eel wall and \SI{90}{\degree} (or \SI{270}{\degree}) for a Bloch wall. In order to fit $\psi$ and $\lambda$, we take the finite resolution of the microscope into account, by assuming a Gaussian beam profile with a standard deviation of \SI{13}{\nano\meter}. Finally, by simultaneously fitting the profiles $m_{\rm i}(x)$ of the wall in the 70-nm-wide nanowire, we find $\lambda = \SI{11\pm2}{\nano\meter}$ and $\psi = \SI{135\pm5}{\degree}$ which is equivalent to $\psi = \SI{45\pm5}{\degree}$. In the following, for convenience, we will display $\psi$ always as the smallest multiple. A small asymmetry in the measurement of the $m_y$ component is present, corresponding to a rotation of \SI{5\pm1}{\degree} of the secondary electrons, which we attribute to a combination of misalignment of the sample normal with respect to the detector axes and spin rotation through the electron optics.

The results of several fitted walls are shown in Table \ref{tab1}.
\begin{table}[h]
\caption{Results of fitted domain wall profiles}
\begin{tabular}{|c|c|c|}
\hline Nanowire width (nm)  & $\lambda$ (nm) & $\psi$ (\SI{}{\degree}) \\ 
\hline 57  & $9\pm2$ & $5\pm12$ \\ 
\hline 70  & $11\pm2$ & $45\pm5$ \\
\hline 93  & $7\pm2$ & $90\pm10$ \\ 
\hline 300 & $9\pm2$ & $89\pm14$ \\
\hline
\end{tabular} 

\label{tab1}
\end{table}
We observe Bloch walls in nanowires of 93-nm width and above, a N\'eel wall in the 57-nm and an intermediate wall in the 70-nm-wide wire. 

In the 1D wall model, the demagnetizing energy determines whether a Bloch or a N\'eel wall is the lowest energy state. Although surface and volume magnetic charges are arranged in a 2D fashion, overall the demagnetizing energy can be considered as a transverse anisotropy which depends on the nanowire dimensions \cite{jung_2008, koyama_2011}. A consequence of the model is that the azimuthal angle changes discontinuously as a function of the nanowire width and thickness \cite{jung_2008, dejong_2015}, contrary to our experimental findings. The possible existence of a ``mixed wall'' was discussed by Aharoni \cite{aharoni_energy_1966} in the context of in-plane magnetized films and discarded for the 1D case by a rigorous energy minimization of all possible configurations. However, he conjectured that wall types with 2D spin arrangements with lower energy might exist, explaining, for instance, the occurrence of cross-tie walls \cite{torok_1965}.

\begin{figure}[t]
	\begin{center}
		\leavevmode
		\begin{minipage}{0.48\textwidth}
			\centering
			\includegraphics[width=0.98\textwidth]{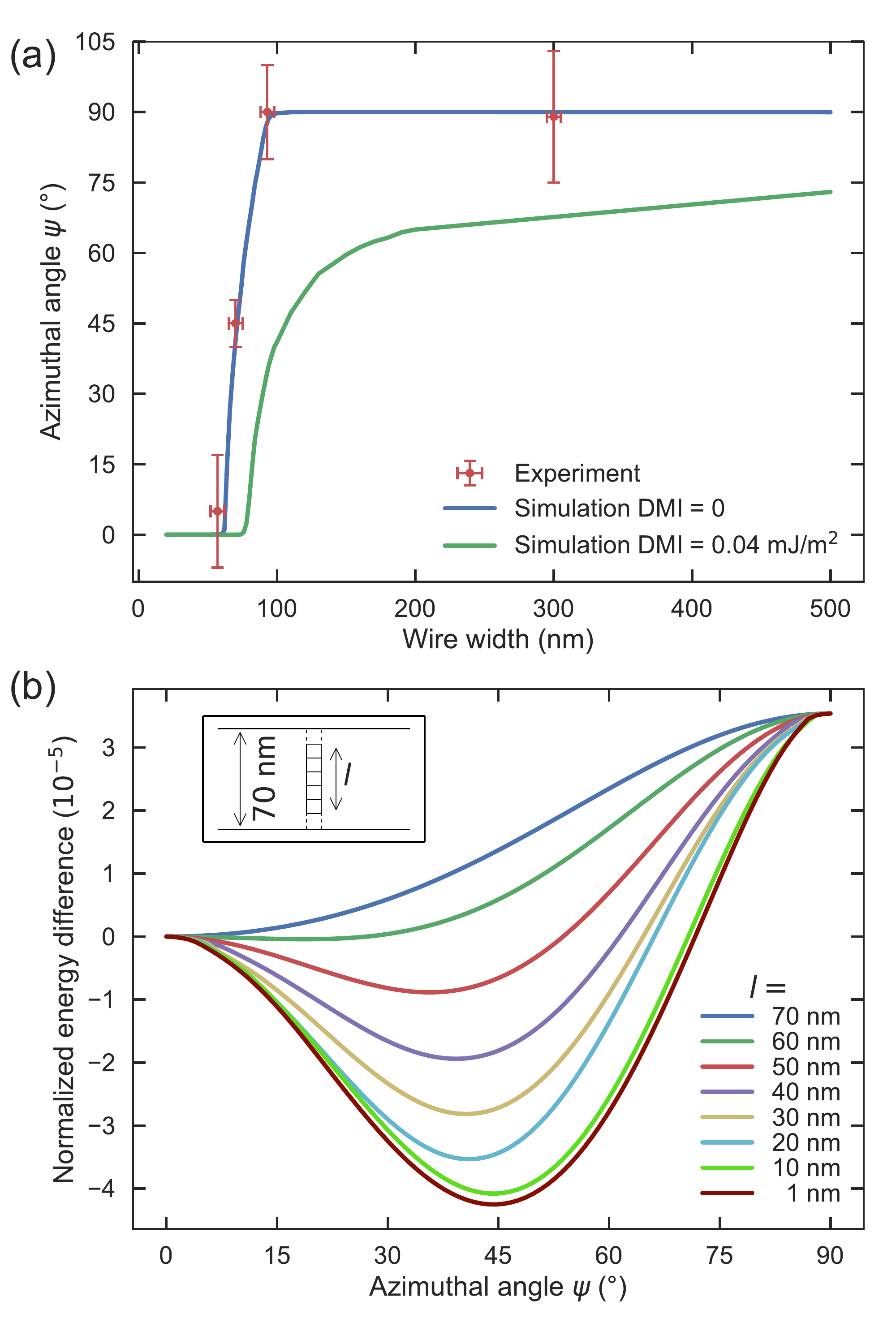}
		\end{minipage}
	\end{center}
  	\vspace*{-5mm}
	\caption{
	(a) Azimuthal angle $\psi$ versus nanowire width extracted from micromagnetic simulations. Without DMI, the domain wall type is N\'eel ($\psi = \SI{0}{\degree}$) for wires up to \SI{60}{\nano\meter}, whereas for wires starting from \SI{98}{\nano\meter}, the type is Bloch ($\psi = \SI{90}{\degree}$). In the transition region $\psi$ changes continuously. By introducing DMI, N\'eel walls are stabilized, shifting the start of the transition to \SI{68}{\nano\meter} and preventing pure Bloch walls even in 500-nm-wide wires. Our measured domain walls from Table \ref{tab1} are included as data points. 
	(b) Normalized energy difference versus $\psi$ for a 70-nm-wide wire. $\psi$ was fixed along a line in the \textit{y}-direction in the center of the domain wall (inset) and varied between \SI{0}{\degree} and \SI{90}{\degree} in steps of \SI{1}{\degree}. The length $l$ of the line of fixed spins varied from \SI{70}{\nano\meter} (entire wire width) down to \SI{1}{\nano\meter} (only cell at center). The energy is plotted as the difference to the N\'eel wall and normalized by the energy of a single domain wire. 
	}
	\label{fig3}
\end{figure}

To overcome the limitations of the 1D model, we performed 2D micromagnetic simulations using OOMMF \cite{oommf}. A wire with a length of 1200 nm and a width varying from 20 to 500 nm was modeled with a cell size of 1 nm $\times$ 1 nm $\times$ 3.8 nm. The material parameters used were $K_{\rm u} = \SI{280}{\kilo\joule\per\cubic\meter}$, $M_{\rm s} = \SI{570}{\kilo\ampere\per\meter}$, and an exchange stiffness $A = \SI{12}{\pico\joule\per\meter}$. For each simulation, the azimuthal angle was deduced from the relaxed energy state, see Fig. \ref{fig3}(a). As expected, we found N\'eel walls for narrow wires and Bloch walls for wide ones. The transition is not abrupt: $\psi$ changes continuously between \SI{60}{\nano\meter} and \SI{98}{\nano\meter}. The width at which this transition occurs depends on the values of the material parameters: For larger $M_s$ or larger $A$ the width gets larger, while for larger $K_u$ it gets smaller.
 

In order to scrutinize the discrepancy between the analytical model and micromagnetic simulations, a series of 2D simulations was run for a wire of 70-nm width. A line of spins within the wall was kept fixed (see inset in Fig. \ref{fig3}(b)), with their azimuthal angle varying from 0 to \SI{90}{\degree} in increments of \SI{1}{\degree}. The energy of the system is plotted in Fig. \ref{fig3}(b). For all spins fixed along the entire width, the lowest energy state is a N\'eel wall; no stable intermediate wall forms, despite the fact that the simulation is 2D. We then sequentially reduced the length of the line of fixed spins. In each series, starting from the edges, more spins were freed until in the extreme case only the spin in the center cell was kept fixed. A pronounced energy minimum develops at a non-trivial angle, i.e., an intermediate wall has formed. This proves that the intermediate wall is a consequence of the 2D nature of a domain wall in perpendicularly magnetized wires.

\begin{figure}[ht]
	\begin{center}
		\leavevmode	
		\begin{minipage}{0.48\textwidth}
			\centering
			\includegraphics[width=0.98\textwidth]{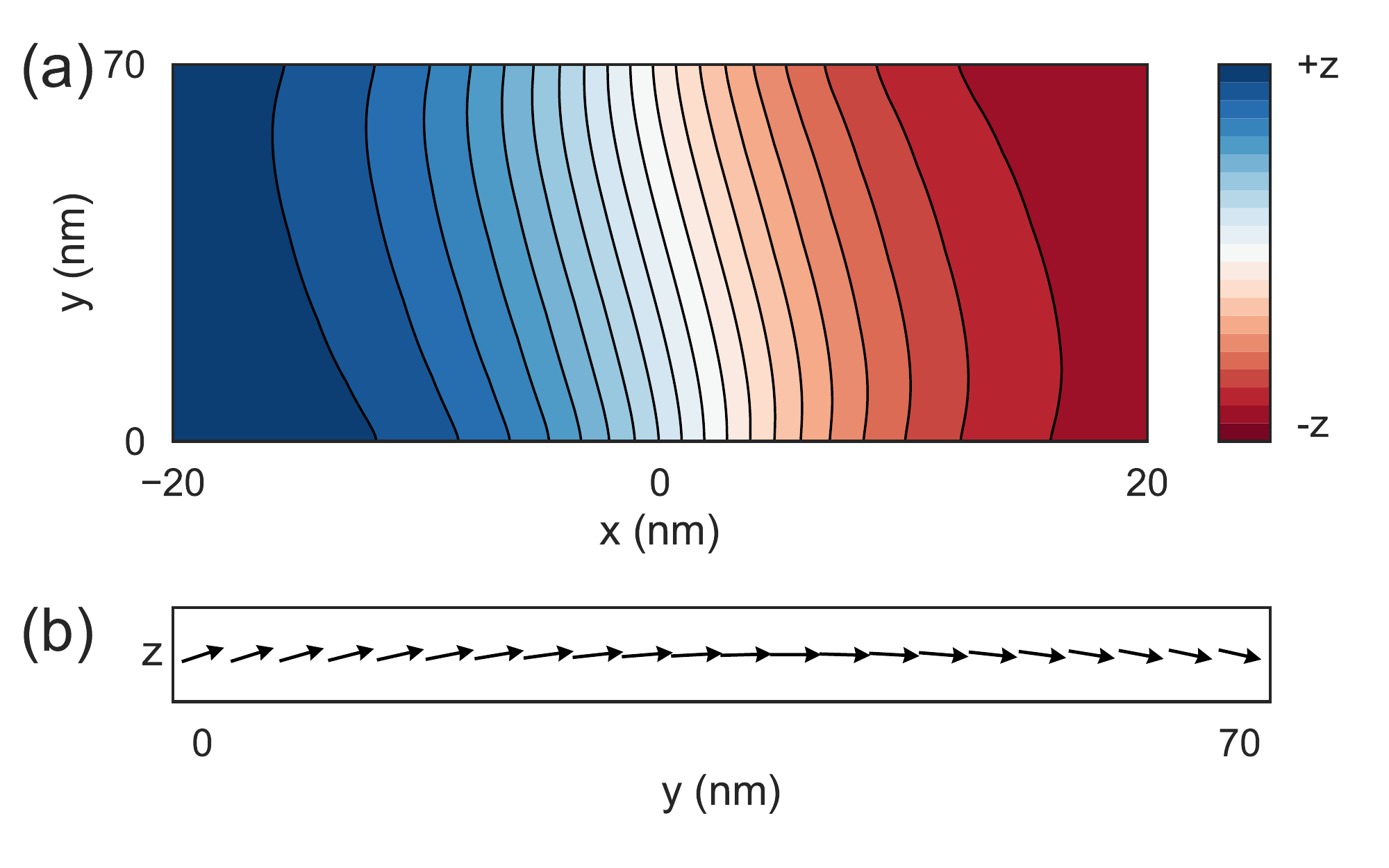}
		\end{minipage}		

 	\end{center}
  	\vspace*{-5mm}
	\caption{
	(a) Top view of a simulated intermediate domain wall in a 70-nm-wide perpendicularly magnetized wire. The out-of-plane component is indicated by color graduation from blue (+\textit{z}) to red (-\textit{z}). The domain wall is slightly wider in the center than at the edges of the wire.
	(b) Cross-section view at $x=0$. The corresponding magnetization direction is indicated with arrows. At the wire edges, a rotation of the \textit{z}-component of the magnetization is observed. It is opposite for opposite edges, leading to a slight canting and an S-shape appearance of the wall. Overall, the wall is inclined by \SI{\sim 3}{\degree} with respect to the wire's cross section.
	Since no chiral interaction is involved an inclination in the opposite direction is equally well possible, provided that the magnetization tilt is also mirrored at the yz-plane.}
	\label{fig4}
\end{figure}

In this 2D wall, the spins tend to align parallel to the edges in order to lower magnetostatic energy, similar to the formation of a N\'eel cap at the surface of a bulk ferromagnet \cite{scheinfein_influence_1989}, and contrary to a tilted 1D wall. In a perpendicularly magnetized ferromagnet, uniaxial anisotropy energy can be gained by tilting the spins within the wall out of the plane. In in-plane magnetized materials, such a tilting generally occurs by forming a C- or S-shaped spin arrangement \cite{guslienko_2001}, with a slight preference for the C-type because of the more complete flux closure of the stray field. Correspondingly, in our perpendicularly magnetized wire, a C-shaped arrangement is set up along the wire's cross section, as shown in Fig. \ref{fig4}(b). Within the wire plane, a C-shape cannot evolve into the adjacent up/down magnetization in a continuous way, and hence an S-shape establishes itself, see Fig. \ref{fig4}(a), with an overall canting angle of the wall of \SI{\sim 3}{\degree}. This inclination of the domain wall can also be seen in the measurements of Fig. \ref{fig1}(b)--(c). The contour levels in Fig. \ref{fig4}(a) reveal that the domain wall is slightly narrower at the edges than in the center, illustrating that spins tilt out-of-plane, which reduces the uniaxial anisotropy energy near the edges. Indeed, the overall energy gain of the intermediate wall compared with the N\'eel wall shown in Fig. \ref{fig3}(b) can be attributed to the anisotropy energy contribution. The azimuthal tilting of the spin within the wall is thus a consequence of the subtle interplay between anisotropy, exchange and demagnetizing energy. The first tilts the spins out of the plane, while the second keeps neighboring spins as aligned as possible. The third one balances the surface magnetic charges of the Bloch wall with the volume magnetic charges of a N\'eel wall. We suspect that an analogous situation exists in in-plane magnetized structures that are too small to support the wide extension of a cross-tie wall. Then the curved and tilted walls proposed long ago \cite{shtrikman_1960} might form.

So far, we neglected another mechanism that can strongly influence the structure of the domain wall: The Dzyaloshinskii--Moriya exchange interaction \cite{heide_2008, thiaville_2012, chen_2013, hrabec_2014}. Its strength is described by the constant $D$, and its energy contribution per unit area for a N\'eel wall is $\pm \pi D$ depending on the chirality of the wall, and zero for the Bloch wall \cite{thiaville_2012}. Strong DMI has been observed in asymmetric Co/Ni multilayers \cite{chen_tailoring_2013}. It influences the Bloch--N\'eel transition by expanding the N\'eel wall regime towards wider and thicker nanowires, so that in films the preferred domain wall will be of N\'eel type or Bloch type with a strong N\'eel component, i.e., a chiral intermediate wall.

In our wires, however, DMI is not the cause of the intermediate domain wall. First of all, we find Bloch  walls
in extended square structures (\SI{70}{\micro\meter} $\times$ \SI{70}{\micro\meter}) and in the film. Second, from Table \ref{tab1}, we see that $\psi = \SI{90\pm10}{\degree}$ in a 93-nm-wide nanowire and $\psi = \SI{89\pm14}{\degree}$ in a 300-nm-wide one. Within the experimental uncertainty, these walls can be considered as Bloch walls with vanishing (or very small) N\'eel component. A N\'eel component induced by DMI would be considerably more pronounced. To substantiate this, the micromagnetic simulations were repeated for wires with $D = \SI{0.04}{\milli\joule\per\square\meter}$. The results are shown in Fig. \ref{fig3}(a). The striking difference is that $\psi$ deviates strongly from \SI{90}{\degree} even at very large nanowire widths, for instance $\psi= \SI{74}{\degree}$ for a width of 500 nm. For the curve shown in Fig. \ref{fig3}(a), we have deliberately chosen a very small value of $D$. The trend to favor a N\'eel wall component is even more pronounced for larger $D$, which is reported in material stacks similar to ours \cite{chen_tailoring_2013}. Therefore, we exclude that the intermediate walls we observe are caused by DMI.

An overall inclination of the wall, which in our intermediate wall is a direct consequence of the magnetization tilt, was observed in domain-wall motion experiments \cite{YamanouchiPRL06, RyuAPE12}. It affects current-induced wall motion because of the induced wall pressure \cite{ViretPRB}. It was proposed that this inclination can be exploited to deduce the DMI value \cite{BoullePRL13}. With our finding of an intermediate wall, one needs to carefully examine in each case whether such an inclination is caused by DMI alone or whether an achiral intermediate wall -- unrelated to DMI -- contributes.

It is remarkable that this new wall type has been overlooked for so long. In wires in which both Bloch and N\'eel walls were identified \cite{koyama_2011}, intermediate walls should show up with a distinct AMR, provided the equilibrium state is attained. In micromagnetic simulations, the intermediate wall is missed if the starting configuration is a Bloch or a N\'eel wall, as for instance in Ref. \cite{martinez_2011}: The energy landscape is too flat there. Analytical approaches \cite{dejong_2015} captured the transition width accurately by developing sophisticated models for the magnetostatic energy, but were also unaware of the existence of a lower-energy 2D wall structure.

In conclusion, we determined the structure of domain walls as a function of the width of perpendicularly magnetized Co/Ni nanowires. Bloch walls prevail for wires wider than \SI{90}{\nano\meter}, N\'eel walls for wires narrower than \SI{60}{\nano\meter}. The transition is not abrupt, contrary to expectations based on the commonly considered 1D model: Intermediate walls form. We showed that such a transition does not require additional effective transverse fields nor DMI. The subtle balance of the various energy terms requires that the magnetization configuration adopts a 2D distribution across the wire. In particular, the spins within the wall tilt out of the plane when approaching the wire edge, in striking contrast to both a Bloch and a N\'eel wall. We argue that this intermediate wall type is a general phenomenon that should occur in any perpendicularly magnetized material provided the wire width is chosen appropriately. This width can be tuned by the perpendicular anisotropy, saturation magnetization, and exchange stiffness.

It would be interesting to investigate the consequences such continuous transition regions have on effects that rely on the discrete Bloch-to-N\'eel transition, such as the reported drastic reduction of the critical current in spin-transfer-driven domain wall motion \cite{jung_2008} or the deferral of the Walker breakdown to higher fields \cite{martinez_2011}.


The authors are grateful to Karen~L.~Livesey for detailed insight into the one dimensional model and Gianfranco Durin and Simone Moretti for discussions. The research leading to these results has received funding from the European Union Seventh Framework Programme [FP7-People-2012-ITN] under grant agreements 316657 (SpinIcur) and 608031 (Wall).


\newpage


\clearpage


\end{document}